\documentstyle[12pt,world_sci]{article}
\begin{document}
\def\Dslash{\,\raise.15ex\hbox{/}\mkern-13.5mu D} 
\title{CURRENT DIRECTIONS IN STRING THEORY AND ABSTRACT SUPERSYMMETRY}
\author{JEFFREY A.  HARVEY \\
{\em Enrico Fermi Institute and Department of Physics,
University of Chicago\\
5640 S. Ellis Ave., Chicago, IL 60637, USA}}
\maketitle
\setlength{\baselineskip}{2.6ex}
\vspace{-1.7in}
\rightline{EFI 94-58}
\rightline{hep-th/9411097}
\rightline{November 1994}
\vspace{1.2in}

\begin{abstract}
\small
I discuss some current thoughts on how low-energy measurements and
consistency constrain theories at high energies with emphasis on string theory.
I also discuss some  recent work on the dynamics of supersymmetric
gauge theories.
\end{abstract}

\section{Introduction}

In her  talk  at this meeting Angela Olinto used both
broad and fine brush strokes to paint her view of  cosmology in
the next millennium.
 In this talk I intend to use only the broadest of brushes, perhaps spray
painting
would be a better analogy, to discuss our ability to probe
Planck scale physics using only low-energy experiments  and self-consistency.
I will also discuss in very general terms some  exciting recent results on
the dynamics of supersymmetric gauge theories.

There is a great deal of doom and gloom in particle physics at the moment
as a result of the cancellation of the SSC and the meager job market for
young particle physicists. There is also great frustration at the success of
the Standard Model and our seeming inability to probe what lies beyond it.
This has led to statements in the popular press implying that particle physics
is approaching a dead end as a field of scientific inquiry. While there is
certainly cause for concern,  I think such pronouncements are  short-sighted
and seriously underestimate the ingenuity  of both
theorists and experimentalists over the long term.  To counter this I would
like
to discuss some reasons for optimism and some new developments which show
that  our bag of tricks is not yet exhausted.

One  valid concern about current particle theory is the vast chasm which
has opened up between experiment and some of the theoretical frontiers,
especially string theory.  There is a tendency in many quarters to regard
string theory as a failure, not because of any internal problems, but because
it only addresses physics at the Planck scale and hence
 is inherently untestable in the foreseeable future.  While direct tests are
certainly
hard to imagine, I would like to discuss a number of indirect ways in which
physics at the Planck scale and particularly string theory are already
playing a significant role in our view of low-energy physics.

Before beginning let me mention a remark which I have heard
with varying attributions to the effect that the problem is not that we take
our theories too seriously, but rather that we do not take them seriously
enough.
Examples abound, ranging from black holes  to quarks to gauge theories
and electroweak theory. In each case the theory was initially viewed
as either a mathematical curiosity or abstraction or as a toy model.
In each case the theory was more real  than physicists could bring themselves
to believe.  As a mature field, particle physics is more tightly constrained
than
newer fields and often a small amount of data is sufficient to develop a
very rigid theory.  Electroweak theory is a particularly good example of this.
I would like to encourage you to take string theory as seriously as I think
it deserves to be taken, not as a mathematical abstraction or a toy model
of quantum gravity, but rather as a real theory of the world that may shed
light
on real phenomena.

\section{Large-scale small-scale connections in particle physics}

\subsection{The Standard Model}
We are currently both blessed and cursed by the incredible success of the
Standard Model
as reviewed at this meeting by Jon Rosner \cite{Rosner}.  In spite of its
success, it is certainly true that it is not a final theory of everything, but
rather
only a low-energy effective theory of  something, namely the light   particles
which include
the three generations of quarks and leptons,
the gauge bosons of $G = SU(3) \times SU(2) \times U(1)$, and presumably
one or more Higgs bosons.  There are many reasons
to believe that the Standard Model is only a stepping stone on the
way to a more complete theory. First of all, it does not include gravity.
Second, it almost certainly
does not exist formally as a quantum field theory due to the lack of asymptotic
freedom of some of the coupling constants. Finally, it has too many loose
ends, requiring the specification of some $20$ odd parameters.  One of the
main goals of particle physics at the end of the $20$th century is to figure
out what lies beyond the Standard Model.

Since we do not believe the Standard Model is a final, self-contained theory,
we should view it only as a low-energy effective theory that  describes physics
in an approximate (but nonetheless remarkable accurate) way at low-energies.
According to the philosophy of effective field theory \cite{JoeP}
 if the Standard Model is effective below a scale $\Lambda$ then
in the effective Lagrangian
we should write down all interactions involving the light degrees of freedom
which are compatible with the symmetries of the problem. Furthermore,
all dimensionfull couplings should  be given by appropriate powers of
$\Lambda$ up to factors which are (roughly) of order one.  Very schematically
this gives
\begin{equation}
{\cal L}_{eff} = ( \Lambda^4 + \Lambda^2  \phi^2)   + ((D \phi)^2 + \bar \psi
\Dslash \psi
+ F_{\mu \nu}F^{\mu \nu} + \phi \bar \psi \psi + \phi^4 ) +  ({\bar \psi \psi
\bar \psi \psi \over
\Lambda^2} + \cdots )
\end{equation}
in a notation where $\phi$ stands for generic scalar field, $\psi$ for generic
fermion fields, and $F_{\mu \nu}$ for the gauge fields.
The first set of terms have positive powers
of $ \Lambda$, they are the most relevant terms in understanding the
low-energy theory.  They consist of the cosmological constant term
and a mass term for any scalar particles.  Note that a mass term for
fermions is forbidden in the Standard Model  by gauge invariance.
The second set of terms are independent of $ \Lambda$
(or have at most log dependence on $ \Lambda$ ) and consist of all
the interactions of the usual Standard Model except for the Higgs
mass term. The final set of terms is really an infinite expansion which
includes all terms with negative powers of $\Lambda$. These terms
are non-renormalizable in the usual sense, and are irrelevant in
very low-energy processes, but are nonetheless there
in a generic low-energy effective field theory.

Now one beautiful thing about the standard model is that {\it all} the
renormalizable terms which are consistent with the gauge symmetries
are present. It is true that some of the couplings are small for poorly
understood reasons (e.g. the electron and  up and down quark Yukawa couplings),
but we are not forced to set any couplings to zero in order to agree
with experiment (except perhaps the cosmological constant ) .

The natural question to ask is, what is the value of $\Lambda$?
Historically the first answer was that $\Lambda = \Lambda_{GUT} \sim
10^{15} GeV$ based on unification of the gauge couplings as well
as on the absence of proton decay, flavor changing neutral currents,
and other  non-standard model processes. The point being that such effects
are generally contained in the non-renormalizable interactions  but
are sufficiently small if suppressed by powers of $\Lambda_{GUT}$.
This answer leads to two great embarrassments which involve
the two relevant terms in ${\cal L}_{eff}$. The first is the
cosmological constant problem. The value of the cosmological constant
is {\it much} less than $\Lambda_{GUT}^4$ for reasons that are completely
mysterious.
The relevance of this term is clear. If the cosmological constant were
of order $\Lambda_{GUT}^4$ the universe would have come and gone
in  a Planck time without any physicists to worry about the problem.
The second  embarrassment is usually called the hierarchy problem, namely that
the
Higgs mass is not $\Lambda_{GUT}^2$ but must be more of order
$M_W^2$. There is of yet no good answer to the first problem,
but at least we can appeal to an ignorance of quantum gravity.
The second problem is a problem of particle physics and most be
faced up to.

If we are not allowed  unnatural fine tuning  then the
only possibility is that $\Lambda$ is not $\Lambda_{GUT}$ but
rather is  not far from the weak scale,
say generically $\Lambda \sim 1 TeV$.  There are currently two main
speculations
as to the new physics at this scale. The first is that there are new strong
interactions,
that is technicolor, top quark bound states or whatever; the second is
that this is the scale of new supersymmetric particles.  Whatever
the new physics, we know that it must have special features. In particular,
we know that many of the non-renormalizable interactions in
${\cal L}_{eff}$ cannot be present if they are suppressed only by powers
of $\Lambda \sim 1 TeV$.  Of course this is not a surprise. It is well
known that technicolor and supersymmetric models are strongly
constrained by the demands that proton decay, CP violation and flavor
changing neutral currents be compatible with present experimental
limits.  But it is worth emphasizing that this is a positive feature. It tells
us that the new dynamics at the TeV scale is not generic but must have
special structures or symmetries which forbid these effects.

\subsection{The MSSM}
Currently the most popular proposal for going beyond the standard model
is based on supersymmetry.  There is now a standard minimal
model incorporating supersymmetry, the MSSM. \cite{howie}
We should first understand how supersymmetry
solves the naturalness problem.  In the early days of supersymmetry one
sometimes heard the following idea. Supersymmetry if unbroken requires
the equality of fermion and boson masses. Since fermions are required to
be massless in the Standard Model before weak symmetry breaking,
bosons would also have to be massless before electroweak breaking.
A Higgs mass of order the weak scale would then be generated provided
that supersymmetry was effectively broken near the weak scale.

This
idea does not work in the most naive sense. The reason is as follows.
In $N=1$ supersymmetry chiral fermions are paired with complex scalars.
Thus one standard Higgs doublet is paired with a chiral fermion doublet.
Now adding such a multiplet to the Standard Model is not possible without
upsetting the delicate cancellation of anomalies between chiral fermions.
So what is done is to add two multiplets corresponding to two Higgs
doublets $H_1, H_2$ with opposite hypercharge. Their fermion
partners are then non-chiral and there is no problem with anomalies.
But then there is a supersymmetric and gauge invariant mass term allowed
for the multiplets which in a supersymmetric notation takes the form
\begin{equation}
W = \mu H_1 H_2 .
\end{equation}
Since this term is supersymmetric and gauge invariant there is no reason
for it not to be of order the Planck scale or whatever scale comes beyond
the MSSM and thus the hierarchy or naturalness problem is not really
solved.  In the context of the MSSM this is usually called the
`` $\mu$ problem ''.

It is tempting to try to forbid this term by some symmetry, but this also
leads to problems, in particular with $\mu=0$ the MSSM has a standard
Peccei-Quinn axion which is ruled out experimentally. So we need
$\mu$ to be non-zero and of order the weak scale.  This suggests
that the value of $\mu$ should be connected with supersymmetry
breaking, even though it is by itself supersymmetric.

There are in fact solutions to the $\mu$ problem in supergravity and
string theory where this is precisely what happens. $\mu$ is zero
initially for reasons that have to do with the precise high-energy
structure of the theory and a non-zero value is induced only
through supersymmetry breaking effects. \cite{muprob}
Thus at least in one class of models the consistency
of the  MSSM requires  the existence of special features
at the Planck scale.

It was argued earlier that the theory beyond the standard model
would have to have special features in order to ensure
that baryon number violation and flavor changing neutral
currents (FCNC) are sufficiently small.  In the MSSM one
can say quite specifically what is required.
The presence of scalar partners of quarks (squarks) means
that there are renormalizable couplings which violate baryon number.
These must be very small to be phenomenologically acceptable.
The simplest solution is to use a discrete symmetry to  force
them to be zero. This is conventionally done by defining a $Z_2$
discrete symmetry called  $R$-parity which has eigenvalue
$(-1)^{3(B-L)+2S}$ when acting on a state of baryon number $B$, lepton
number $L$ and spin $S$. Equivalently, it is $+1$ on all standard model
particles
and $-1$ on all their superpartners.
$R$ parity also forbids all renormalizable
couplings which violate lepton number. $R$ parity also implies that
the lightest supersymmetric partner (LSP) is absolutely stable since there
is nothing it can decay into while preserving $R$-parity.
Thus $R$-parity has very important phenomenological consequences.
There are other
discrete symmetries which do not forbid all these couplings
but which nonetheless are phenomenologically acceptable.
Perhaps the most attractive is a $Z_3$ matter parity discussed
by Ibanez and Ross. \cite{IbRoss} The central position played by such
discrete symmetries makes it important to ask whether we really expect
exact discrete symmetries to exist.

Global discrete symmetries may be approximate symmetries
of low-energy effective Lagrangians, but it is difficult to see
why they should not be violated at sufficiently high energies,
say by gravitational effects.  However discrete symmetries can
be exact if they are gauge symmetries. The distinction between
gauge and global symmetries applies whether the symmetry is
continuous or discrete. Global symmetries are true symmetries
of the configuration space of a theory while gauge symmetries
are only a redundancy in our description of the theory or equivalently
of its configuration space.  This is why one does not find Nambu-Goldstone
bosons
in ``spontaneously broken'' gauge theories.  Gauge symmetries
cannot be spontaneously broken since they are just a redundancy
of our description,  equivalently there are no flat directions in
the configuration space due to gauge symmetries which could
give rise to massless modes.

With this in mind it should be clear that discrete symmetries can also
be classified in the same way.  One simple way such symmetries can arise
is by ``breaking''  of a $U(1)$ gauge symmetry down to a $Z_N$ subgroup.
The low-energy theory will have a $Z_N$  discrete symmetry which is
a gauge symmetry since it is a subgroup of the original $U(1)$ gauge
symmetry.  Such discrete gauge symmetries can arise by other means
and in fact they are very common in compactifications of string theory.
There are also consistency conditions for discrete gauge symmetries
coming from a discrete version of anomaly cancellation, \cite{discanom}
these conditions are satisfied
both for the usual $Z_2$ R-parity and for the $Z_3$ matter parity.
Thus we again find that a phenomenological requirement on the
MSSM leads us to conclude that it should be imbedded in a theory
at high energies which can contain  discrete gauge symmetries.

Probably the outstanding problem in the MSSM is understanding
the mechanism responsible for supersymmetry breaking.  Ultimately the
hierarchy problem must be solved by eplaining why the effective scale
of supersymmetry breaking is near the weak scale.  Also, without
an understanding of supersymmetry breaking   we can only
parametrize the breaking  by soft terms in the low-energy Lagrangian.
Although only four such terms are often used based on
certain ``minimal'' assumptions, these assumptions are
totally unjustified from a theoretical point of view and
there are really $60$  odd parameters needed to specify
the most general soft breaking terms.  As a result the
model has little predictive power.  At present we don't know whether
supersymmetry is broken by the dynamics of some hidden
strongly coupled gauge sector, by non-perturbative effects
in the visible sector, or by some intrinsically stringy
mechanism. Up until recently the most popular models have
been hidden sector models where supersymmetry is actually
broken at a rather large scale, but only communicated to the
visible world by gravitational effects, leading to an effective scale of
supersymmetry breaking around the TeV scale.

Thus in the MSSM there are many hints as to  the structure of whatever
theory lies beyond the MSSM, perhaps at the Planck scale.  It must
solve the $\mu$ problem, allow for discrete symmetries, preferably gauged,
and perhaps provide a mechanism for dynamical supersymmetry breaking.
Such a mechanism is further constrained by the allowed values of the soft
supersymmetry breaking parameters.

\subsection{String Theory}

In the previous two sections I have tried to argue that the low-energy
structure
we have observed and hope to observe in the future carries important clues
about the behavior of physics at very short distances, e.g. the Planck scale.
Currently the most successful model of Planck scale physics is superstring
theory.
It is not my intention to review string theory or even current progress in
string
theory in this talk. Instead I would just like to mention a rather simple but
important way in which string theory itself provides a connection between
physics on vastly different scales.

This comes about by asking the question `` How big is a string?''.
Consider for simplicity a closed bosonic string which traces out
a closed loop in space parametrized by a coordinate $\sigma$ with
$0 < \sigma \le 2 \pi$.  We can write a normal mode expansion
 for the coordinate of the string as

\begin{equation}
X(\sigma) = x_{cm} + \sum_{n} \left( {x_n \over n} e^{i n \sigma} + {\tilde x
\over n}
                                                                        e^{-i n
\sigma} \right)
\end{equation}
with $x_{cm}$ the center of mass of the string and $x_n$, $\tilde x_n$ being
the
Fourier coefficients. When time dependence is added the $x_n$ correspond to
excitations running around the string counterclockwise and the $\tilde x_n$ to
excitations running around the string clockwise.

One measure of the size of the string is the average deviation of any point
on the string from the location of the center of mass, that is
\begin{equation}
R^2 = \langle (X(\sigma) - x_{cm})^2 \rangle
\end{equation}

We want to calculate this average quantum mechanically in which case
the $x_n$ and $\tilde x_n$ become harmonic oscillator creation and
annihilation operators for  $n<0$ or $n>0$ respectively.  Clearly for a very
excited string $R$ will be very large, but what is rather strange is that due
to zero-point fluctuations $R$ is large even in the string ground state.
In the string ground state one easily finds that  the $n^{th}$ oscillator
contributes
a factor of $1/n$ and
\begin{equation}
R^2 \sim \sum_{n} {1 \over n}
\end{equation}

To make sense of this divergent sum we have to understand what it means
physically to measure the size of a string. Any real measuring device will
have some finite time resolution $\tau_{res}$. On the other hand the divergence
comes from modes with large $n$ and these modes have frequencies which
grow as $n$.  Since a real measuring device cannot  measure  frequencies
greater than $1/\tau_{res}$ it makes sense to cut off the sum at a mode number
corresponding to the time resolution of the measuring device. Doing this
somewhat more carefully than described here and reinstating dimensions
gives \cite{Sussa}
\begin{equation}
R^2 \sim l_s^2 \log ( P_{tot}/\tau_{res} )
\end{equation}
where $l_s$ is the string scale which is of order the Planck scale (it is less
than the Planck scale by a power of the dimensionless string coupling
constant) and $P_{tot}$ is the total transverse momentum of the string.

This equation is quite remarkable when one thinks about it. It says that
the size of a string depends on how fast you can measure it, and that
the faster you can measure it, the bigger it appears.  We are used to the
idea, which follows from the uncertainty principle, that we need high energies
(short  times) to probe short distance scales. But as we approach the Planck
scale things change, at least in string theory. We need high energies (short
times)
to probe large distance scales, or at least the large size behavior of strings.

Considerations like this lead to a ``string uncertainty principle''
\cite{grossetc}
which
relates the uncertainty in the size of a string $\Delta x$ to its
energy $E$:  (in units with $\hbar = c=1$)
\begin{equation}
\Delta x \ge {1 \over E} + {E \over M_{Pl}^2 }.
\end{equation}
This equation incorporates
both the usual quantum mechanical uncertainty and the less well understood
uncertainty due to the growth of string states at high energies.

This behavior is responsible for many qualitative features of string theory.
For example it is generally believed that string theory has no ultraviolet
divergences and this is certainly backed up by many explicit calculations.
The above behavior explains why.  At very high energies strings become
very large and floppy and there is no large concentration of energy
at a point which
could lead to bad high-energy behavior.
It also sounds temptingly like what one wants in order to solve the
cosmological
constant problem. There one needs some peculiar connection between
large scale physics (as governed by the smallness of the cosmological
constant which explains the largeness of the present universe) and small
scale physics (which should govern the leading contributions to the
cosmological constant. ) However I do not know of any serious attempt to
solve this problem in string theory which utilizes this feature of string
theory.

\subsection{Black Holes}

Over the last few years there has been a resurgence of interest in the
quantum  mechanics of black holes, inspired in part by the hope that
string theory or tools derived from string theory may help to solve some of the
outstanding puzzles.  The basic outstanding questions are

\begin{enumerate}

\item What is the correct treatment of the black hole singularity?  Is the
breakdown of
classical general relativity a sign that quantum gravity is important,  or that
new short distance classical effects are present (e.g. classical string
theory)?

\item Is the Hawking evaporation of black holes consistent with the formalism
of quantum mechanics? In particular, can pure states evolve to mixed states
due to black hole intermediate states?

\item Is there a microscopic description in terms of state counting of the
Bekenstein-Hawking
black hole entropy $S= A/4$ with $A$ the area of the event horizon?

\end{enumerate}

In spite of much activity, there is as yet no consensus on the answers to these
questions.  If black hole entropy is  associated with  new states on the
horizon
of a black hole then
these states must either show up in a more careful treatment of quantum gravity
in the presence of black holes or they  must be associated to new states in
a more complete theory of quantum gravity such as string theory. The first
point of view has been explored recently in the context of toy models
in $2+1$ dimensions by Carlip, Teitelboim and collaborators \cite{entrop}.  The
second
point of view has been explored by Susskind and collaborators \cite{bhrefs}
and fits in nicely with
 the behavior of string theory discussed earlier and with
the general idea of looking for connections between short and long distance
physics.

Following Susskind and collaborators consider  the difference between
a point particle and a fundamental string falling into a black hole as seen
by an outside observer. In either case, the outside observer sees any
radiation or signal coming from the infalling object redshift  exponentially
as it approaches
the horizon. Thus a measurement made at a fixed frequency probes the structure
of the infalling object on increasingly shorter time scales. According to the
previous
description this should mean that the outside observer in effect ``sees'' a
string
increase in size as it falls into a black hole. This suggests that as it
reaches the
horizon the string is effectively smeared over the whole horizon of the black
hole.
It seems plausible that such smeared string states could account for the
entropy of the
black hole, but there has so far been no explicit calculation which shows this
to be the case. Also, the word ``see'' should be taken with a grain of salt.
Because
of the exponential redshift in the the radiation emitted by an object falling
into
a black hole, it becomes harder and harder to actually  see the physics of
the string spreading. Presumably it eventually becomes mixed up with the
Hawking radiation in some complicated way.

The question of loss of coherence or information in black hole formation and
evaporation should in some sense be tied to the microscopic description of
entropy. If it is possible to account for the states and their thermal behavior
in an honest way, then we would expect that there would be no information
loss once the detailed interaction with the horizon states is take into
account.
On the other hand, the whole notion of new states on the horizon is rather
peculiar and contrary to many of the developments in general relativity over
the last twenty years. In particular, to an  infalling observer there is
absolutely
nothing special about the horizon and no reason to expect any new states,
nor anyplace they would be expected to be. Thus there would have to be some
very strange new kind of complementarity between the description of physics
by different observers.

There have also been many attempt to resolve these problems in toy models
based on $1+1$ dimensional versions of gravity, but again there seems to
be no consensus.  An up to date overview of this problem can now be
found on the Web. \cite{giddings}

\subsection{Inflation}
I will be very brief here since Inflation has already been discussed at this
conference
in the lecture by Angela Olinto. \cite{olinto} There are several reasons why
inflation
provides a particularly compelling example of  relations between small scale
and large scale physics.  First of all, there is so far no microscopically
compelling model
of inflation, that is  a natural particle physics model which gives rise
to an inflaton field and potential that leads to the proper amount of inflation
and the density perturbations of the required amplitude.  Such a model would
provide a direct link between physics near the GUT scale and the size of
the current universe.  More spectacularly,
in inflation models the density perturbations which eventually grow into
the stuff we see about us are supposed to have their origin in the quantum
mechanical fluctuations of the inflaton field, what more dramatic connection
between microscopic and macroscopic physics could we hope for? Finally,
our current understanding of inflation rests on the idea that there was a past
epoch of the universe where the vacuum energy density, a.k.a. the cosmological
constant was non-zero. While this seems inescapable in many models given
that the present cosmological constant is zero, one cannot escape an uneasy
feeling that our whole picture of inflation may change dramatically if we ever
understand the cosmological constant problem.

\section{Dynamics of Supersymmetric Gauge Theories}

\subsection{Motivation}
During the last $15$ years two theories have been developed in some detail that
involve supersymmetry in an essential way. The first is superstring theory.
Superstring
theory allows us for the first time to address the physics of the Planck scale
in a well
defined way. It is also well understood by now that there are many solutions to
superstring theory that leave us with an effective four-dimensional
supersymmetric
theory which resembles the standard model in the sense that it can contain
some number of chiral fermion generations in representations of a gauge group
which is usually somewhat larger than the gauge group of the standard model.
One the other hand there is so far no understanding of which (if any) of these
vacua is picked out dynamically as the true vacuum (they are equally good in
perturbation theory).  There may also be many inequivalent vacua left after
all dynamical effects are included in which case one needs some further
principle
(e.g. quantum cosmology) in order to make contact with our particular world.
Another aspect of  this problem is that most superstring vacua contain many
massless
scalar
fields called moduli which have no potential. Different vacuum expectation
values
for these moduli correspond to different choices of superstring vacua.  The
``moduli problem''
in superstring theory is the problem of how a suitable potential is chosen by
the
dynamics for these fields.

The second theory where supersymmetry has played a crucial role
is of course the supersymmetric extension of the standard model (MSSM)
discussed earlier. There the main roadblock to detailed predictions is our lack
of understanding of supersymmetry breaking.

One of the main themes in particle theory over the last few years has been the
attempt
to tie together the structure needed in  the MSSM  with the sort of structures
that
arise in superstring theory.  This not only provides further constraints on the
MSSM but also may shed some light on both the moduli problem and the
problem  of supersymmetry breaking.

Over the last year there has been dramatic progress, building on work
in the early 1980's,  in understanding the dynamics
of supersymmetric gauge theories in four dimensions. Although this work has not
yet answered either of the above questions, it has introduced new techniques,
and
also promises to shed light on some old problems in non-supersymmetric gauge
theories.
As a result I would like to give a brief discussion of these new developments
following a recent paper by Seiberg and Witten. \cite{SW}

\subsection{$N=1$ Supersymmetry}

Before discussing the results of Seiberg and Witten it will be useful to very
briefly
review a few facts about theories with $N=1$ supersymmetry.  These
theories are the basis for supersymmetric extensions of the standard model
(the MSSM) because only theories with $N=1$ supersymmetry are compatible
with having chiral fermions.  Lagrangians with $N=1$ supersymmetry are most
easily constructed in terms of superfields in superspace. The coordinates
of superspace consists of the usual  spacetime coordinates $x^\mu$ and
fermionic coordinates $\theta$ which can be thought of as the two complex
components of a Weyl fermion.  The basic matter multiplet, consisting of
a  Weyl fermion and a complex boson  can be packaged into a  complex
scalar function on superspace, $\Phi (x,\theta)$ called a chiral superfield.
The general Lagrangian then has two types of terms, written as
\begin{equation}
{\cal L} = ( \int d^2 \theta d^2 {\bar \theta} K ( \Phi, {\Phi}^{*} ) )  + (
\int d^2 \theta  W(\Phi) + h.c.)
\end{equation}
and referred to as $D$ terms and $F$ terms respectively. For a renormalizable
Lagrangian the $D$ terms contain  the kinetic energy terms while the $F$ terms
contain potential  terms and Yukawa interactions.  The important point to note
is that the $F$ terms are determined purely by functions of $\Phi$ while the
$D$
terms involve functions of both $\Phi$ and $\Phi^{*}$. This is peculiar to
supersymmetry and has no analog in ordinary non-supersymmetric field theories.
It has been known some time that $F$ terms are not renormalized in perturbation
theory. The proofs of this rely on the detailed structure of superspace
perturbation
theory. More recently it has been realized that there is a very simple and
powerful
argument for this non-renormalization which can also be extended to prove
results about non-perturbative effects.

The argument is best illustrated with a
simple example. \cite{nati} For a single scalar superfield the most general
renormalizable
choice of $W$ is
\begin{equation}
W(\Phi) = { m \over 2}  \Phi^2 + {\lambda \over 3}  \Phi^3
\end{equation}
When $m=\lambda=0$ this theory would have two $U(1)$ symmetries,
consisting of  changing the phase of the scalar and fermion components of
$\Phi$ separately. The idea is to view the parameters $m$ and $\lambda$
as expectation values of chiral superfields and these two symmetries as
being spontaneously broken. Thus we replace
\begin{eqnarray}
 m  & \rightarrow  \langle \Phi_m \rangle \\
                \lambda & \rightarrow \langle \Phi_\lambda \rangle
\end{eqnarray}
and find that the theory is invariant under the two $U(1)$ symmetries
\begin{eqnarray}
S:  & \Phi_m & \rightarrow e^{-2 i \alpha} \Phi_m \\
          &          \Phi_\lambda & \rightarrow e^{-3 i \alpha} \Phi_\lambda \\
           &         \Phi     &     \rightarrow  e^{i \alpha} \Phi
\end{eqnarray}
and
\begin{eqnarray}
R:  & \Phi_m (\theta,x) & \rightarrow  \Phi_m (\theta,x) \\
       &      \Phi_\lambda (\theta,x) & \rightarrow
           e^{- i \beta} \Phi_\lambda (e^{i \beta
                } \theta,x) \\
        &      \Phi (\theta,x) & \rightarrow  e^{i \beta} \Phi (e^{-i \beta}
\theta,x)
\end{eqnarray}
Now since we can view the symmetry as being spontaneously broken,
we know that it must be respected by the one-particle irreducible effective
action. But the only terms allowed in the 1PI effective action by these
two symmetries are the original terms! Thus there can be no renormalization
of the potential terms. It should be mentioned that there are some subtleties
with this argument when there are massless particles. \cite{Dine}
Similar arguments can be used
to determine various non-perturbative effects in these theories, but only
through the contribution to $F$ terms. The argument does not work for
$D$ terms since they depend on both $\Phi$ and $\Phi^{*}$.  If there
were an additional symmetry relating $D$ and $F$ terms then one would
have very strong constraints on the dynamics of the theory. This is precisely
what happens in theories with $N=2$ supersymmetry.

\subsection{$N=2$ Supersymmetry}

The simplest  Yang-Mills theory with $N=2$ supersymmetry contains
a single supermultiplet of fields consisting of what we might
call a gauge boson and gaugino plus a Higgs boson and Higgsino.
These fields are all in the adjoint representation of the gauge group.
Taking the gauge group to be  $SU(2)$ they are all isotriplet fields. The Higgs
$\phi$ is a complex field. $N=2$ supersymmetry dictates that the potential
term for $\phi$ is
\begin{equation}
V(\phi) = {\rm Tr} [\phi, \phi^{*}]^2 .
\end{equation}
Clearly $V=0$ for any configuration of $\phi$ which lies in a single direction
in group space, say $\phi = a \tau_3 /2$ with $a$ an arbitrary complex number.
A gauge invariant way of describing this is to say that there is a zero of
$V$ for every complex value of $u = {\rm Tr} \phi^2 = a^2/2$.

The fact that classically there is a continuous set of vacua specified by
$u$ is a feature which is common to many supersymmetric theories.  In
general it is a disaster for phenomenological applications of supersymmetry.
This is both because it reflects the presence of massless scalars in
the low-energy theory and because it  makes it unclear which vacuum is the
right one for doing physics. This latter problem is particularly onerous in
low-energy supersymmetric string theory.   The beauty of the recent ideas
is to take this bad feature of supersymmetric theories and to make use of
it to study the dynamical structure of these theories.

To see  how this works first consider this theory at some non-zero value
of $u$. We then have $SU(2)$ broken down to a $U(1)$ which I will call
electromagnetism.
The spectrum of the theory consists of a massless photon supermultiplet,
massive
$W^{\pm}$ supermultiplets, and also massive monopole supermultiplets.  At very
low-energies the theory just looks like a supersymmetric form of
electromagnetism
with some electric coupling $e_{eff}$.  I first want to explain how $e_{eff}$
is related
to the value of $u$.  This theory is asymptotically free at high energies, that
is
$\beta_e < 0$ at scales above the masses of all particles. As we drop below the
masses of charged particles they decouple from loops and the coupling constant
stops running. Clearly the asymptotic value of the coupling constant at low
energies
depends on the scale of masses in the theory. These masses are determined by
 the value of $u$.
So, specifying a value of $u$ is equivalent to specifying the coupling
$e_{eff}$ in
the low-energy theory.

As long as we only consider the massive electrically charge particles and the
photon then the low-energy physics is completely specified by $e_{eff}$. When
we include the monopoles as well it is necessary to also specify the effective
value of the $\theta$ angle, $\theta_{eff}$. This parameter is the
electromagnetic
analog of the $\theta$ angle in $QCD$. In QED it also leads to CP violation by
giving magnetic monopoles of  magnetic charge $n_M$ an electric charge given
by
\begin{equation}
Q_{el} = {n_M \over 2 \pi} \theta_{eff}
\end{equation}
The upshot of this is that the complex parameter $u$ in fact determines
two real low-energy parameters $(e_{eff}, \theta_{eff})$ or equivalently,
one complex low-energy parameter.

Now we would like to know what happens to the theory and the spectrum of
states as we move around between classical vacua labelled by $u$, or
equivalently
move in the space of couplings $(e_{eff}, \theta_{eff})$ that govern the
low-energy field theory.  First of all it can be shown fairly easily that at
weak
coupling  $\theta_{eff} \propto {\rm Im} u$. Thus at weak coupling the effect
of shifting the imaginary part of $u$ is simply to shift the electric charge of
magnetic monopole states.  Clearly it would be much more interesting to
determine what happens when $g_{eff}$ changes since this could tell
us about the behavior of the theory at strong coupling.

Unfortunately the analysis get quite a bit more complicated and I cannot
do it justice in this talk. Let me just mention the following points.
\begin{enumerate}
\item The one-loop beta function predicts that the coupling $e^2$ should blow
up in the infrared and then become negative. Clearly this is physically
unacceptable.
Of course there is no reason to believe the one-loop result at strong coupling.
\item The possible behavior of the theory as a function of $u$ is highly
constrained
by the $N=2$ supersymmetry.  In particular the $u$ plane must be a special kind
of complex manifold, a Kahler manifold.
\item By utilizing the supersymmetry and analytic structure one can find a
simple
guess for the behavior of the theory as a function of $u$ which passes many
non-trivial consistency tests.
\end{enumerate}

The resulting structure found by Seiberg and Witten has many remarkable
features.
First of all, there is a kind of ``dual'' behavior at strong coupling where as
one
moves in $u$ the electrically charged states pick up magnetic charge, just
as the magnetic states can pick up charge at weak coupling. Second, one
finds singular points where monopole states are becoming massless.  It is
possible
to perturb the theory by breaking $N=2$ supersymmetry to $N=1$ supersymmetry
in such a way that the monopoles mass squared becomes negative, indicating an
instability to monopole condensation. It has long been thought that confinement
may be described by a ``dual '' superconductor in which magnetic charges
condense. In these models one has for the first time a concrete realization
of these ideas.

There seem to be two major areas where these results may have a broad impact.
The first is in the construction of new models for dynamical supersymmetry
breaking.
As I mentioned earlier, the lack of a convincing mechanism for supersymmetry
breaking is one of the major obstacles to obtaining predictions from the MSSM.
The second area is the dynamics of strongly coupled gauge theories,  both
supersymmetric
and non-supersymmetric.  The models discussed by Seiberg and Witten are  in
a sense toy models, but they contain most of the features of non-trivial gauge
theories including running couplings, non-trivial scattering, confinement, and
chiral symmetry breaking. In addition the way these effects are realized is
rather
dramatic involving an infinite resummation of non-perturbative instanton
effects.
It seems possible that some of these features may extend beyond these specific
models.

\section{General Remarks}

I would like to end this talk with a few general remarks about the prospects
for progress in string theory in the upcoming years.

The string revolution is now ten years old.  The discovery of anomaly
cancellation
by Green and Schwarz was discussed at the November 1984 DPF meeting. Roughly
a year later the basics of superstring models of unification were established
through
Calabi-Yau compactifications of the heterotic string.  Since then there has
been a great
deal of effort and progress in formal areas, but little that an experimentalist
would
find applicable to current or future experiments.  As a result  a certain
amount
of pessimism and scepticism has arisen regarding the future of string theory.

It is worthwhile recalling the situation before string theory.  As is the case
now,
in 1984 there were no startling new experimental results and the leading
idea for unification, Kaluza-Klein theory, had run into insurmountable
obstacles.
However there were a number of novel theoretical ideas and structures floating
around which did not fit into any coherent whole.  As examples I might mention
connections between the renormalization group and geometry,
the beautiful structures of two-dimensional current algebras or
Kac-Moody algebras,  the development of realistic supersymmetric extensions
of the standard model with hidden sector supersymmetry breaking, as well as
a number of other purely theoretical constructions.   Many of these are now
regarded
as either part of the structure underlying string theory or as possible
low-energy
consequences of string theory.

If we think of the situation today there are again many new ideas floating
around,
many are classified in a general way as ``string theory'' but in fact many have
very
little to do with string theory in any direct way.  For example there are
matrix models
of two-dimensional string theory, new ideas concerning the structure of black
holes
with toy models in $1+1$ and $2+1$ dimensions, topological field theory with
its
connection to many deep mathematical structures, new ideas about the structure
and
dynamics  of
supersymmetric gauge theory, a deeper understanding of special two-dimensional
systems, etc.  Of course this is not to say that another revolution
comparable to string theory is around the corner,  but rather that there has
been continuing
theoretical progress and that there are new connections and structures which
will undoubtedly fit into a more coherent picture in the future.

I have tried to take an optimistic, but I hope not unreasonable point of view
in this talk.  My point of view is that in spite of the dearth of new
experimental results, there are new interesting theoretical ideas and tools
and that these plus consistency and constraints from the Standard Model
provide us with non-trivial information about the structure of string theory or
whatever theory governs physics at the Planck scale.  I am certainly not
suggesting that we can find the theory of everything without additional
experimental
input.  I am suggesting that as our theories become more tightly constrained
each additional piece of experimental information carries much more weight.
I think this holds out  hope that we will eventually understand Planck
scale physics without building a Planck scale accelerator.

\section{Acknowledgements}

I would like to thank the organizers for the invitation to speak and the other
speakers for their lively and interesting presentations.

\def \ap#1#2#3{{\it Ann. Phys. (N.Y.)} {\bf#1} (#3) #2}
\def \apny#1#2#3{{\it Ann.~Phys.~(N.Y.)} {\bf#1} (#3) #2}
\def \app#1#2#3{{\it Acta Physica Polonica} {\bf#1} (#3) #2}
\def \arnps#1#2#3{{\it Ann. Rev. Nucl. Part. Sci.} {\bf#1} (#3) #2}
\def \arns#1#2#3{{\it Ann. Rev. Nucl. Sci.} {\bf#1} (#3) #2}
\def \hb87{{\it Proceeding of the 1987 International Symposium on Lepton and
Photon Interactions at High Energies,} Hamburg, 1987, ed. by W. Bartel
and R. R\"uckl (Nucl.~Phys.~B, Proc. Suppl., vol. 3) (North-Holland,
Amsterdam, 1988)}
\def \ib{{\it ibid.}~}
\def \ibj#1#2#3{{\it ibid.} {\bf#1} (#3) #2}
\def \ijmpa#1#2#3{{\it Int.~J. Mod.~Phys.}~A {\bf#1} (#3) #2}
\def \ite{{\it et al.}}
\def \jpb#1#2#3{{\it J. Phys.} B {\bf#1} (#3) #2}
\def \jpg#1#2#3{{\it J. Phys.} G {\bf#1} (#3) #2}
\def \kdvs#1#2#3{{\it Kong.~Danske Vid.~Selsk., Matt-fys.~Medd.} {\bf #1}
(#3) No #2}
\def \mpla #1#2#3{{\it Mod. Phys. Lett.} A {\bf#1} (#3) #2}
\def \nc#1#2#3{{\it Nuovo Cim.} {\bf#1} (#3) #2}
\def \np#1#2#3{{\it Nucl. Phys.} {\bf#1} (#3) #2}
\def \pisma#1#2#3#4{{\it Pis'ma Zh. Eksp. Teor. Fiz.} {\bf#1} (#3) #2 [{\it
JETP Lett.} {\bf#1} (#3) #4]}
\def \pl#1#2#3{{\it Phys. Lett.} {\bf#1} (#3) #2}
\def \plb#1#2#3{{\it Phys. Lett.} B {\bf#1} (#3) #2}
\def \ppnp#1#2#3{{\it Prog. Part. Nucl. Phys.} {\bf#1} (#3) #2}
\def \pr#1#2#3{{\it Phys. Rev.} {\bf#1} (#3) #2}
\def \prd#1#2#3{{\it Phys. Rev.} D {\bf#1} (#3) #2}
\def \prl#1#2#3{{\it Phys. Rev. Lett.} {\bf#1} (#3) #2}
\def \prp#1#2#3{{\it Phys. Rep.} {\bf#1} (#3) #2}
\def \ptp#1#2#3{{\it Prog. Theor. Phys.} {\bf#1} (#3) #2}
\def \rmp#1#2#3{{\it Rev. Mod. Phys.} {\bf#1} (#3) #2}
\def \ufn#1#2#3#4#5#6{{\it Usp.~Fiz.~Nauk} {\bf#1} (#3) #2 [Sov.~Phys. -
Uspekhi {\bf#4} (#6) #5]}
\def \yaf#1#2#3#4{{\it Yad. Fiz.} {\bf#1} (#3) #2 [Sov. J. Nucl. Phys. {\bf #1}
 (#3) #4]}
\def \zhetf#1#2#3#4#5#6{{\it Zh. Eksp. Teor. Fiz.} {\bf #1} (#3) #2 [Sov.
Phys. - JETP {\bf #4} (#6) #5]}
\def \zhetfl#1#2#3#4{{\it Pis'ma Zh. Eksp. Teor. Fiz.} {\bf #1} (#3) #2 [JETP
Letters {\bf #1} (#3) #4]}
\def \zp#1#2#3{{\it Zeit. Phys.} {\bf#1} (#3) #2}
\def \zpc#1#2#3{{\it Zeit. Phys.} C {\bf#1} (#3) #2}

\end{document}